\documentclass[preprint]{aastex}
\usepackage{float,graphicx,amsmath,multirow}
\usepackage{color}

\title{An observational investigation of the identity of B11244 ($l$-C$_3$H$^+$/C$_3$H$^-$)}
\author{Brett A. McGuire \& P. Brandon Carroll}
\affil{Division of Chemistry and Chemical Engineering, California Institute of Technology, Pasadena, CA 91125}
\author{Pierre Gratier}
\affil{IRAM, 300 rue de la Piscine, 38406 Saint Martin d'H\`{e}res, France and LERMA, UMR 8112, CNRS and Observatoire de Paris, 61 avenue de l'Observatoire, 75014 Paris, France}
\author{Viviana Guzm\'{a}n}
\affil{IRAM, 300 rue de la Piscine, 38406 Saint Martin d'H\`{e}res, France}
\author{Jerome Pety}
\affil{IRAM, 300 rue de la Piscine, 38406 Saint Martin d'H\`{e}res, France and LERMA, UMR 8112, CNRS and Observatoire de Paris, 61 avenue de l'Observatoire, 75014 Paris, France}
\author{Evelyne Roueff}
\affil{LUTH, UMR 8102, CNRS and Observatoire de Paris, Place J. Janssen, 92195 Meudon Cedex, France}
\author{Maryvonne Gerin}
\affil{LERMA, UMR 8112, CNRS and Observatoire de Paris, 61 avenue de lÕObservatoire, 75014 Paris, France}
\author{Geoffrey A. Blake}
\affil{Division of Chemistry and Chemical Engineering and Division of Geological and Planetary Sciences, California Institute of Technology, Pasadena, CA 91125}
\author{Anthony J. Remijan}
\affil{National Radio Astronomy Observatory, Charlottesville, VA 22903}

\begin{document}

\begin{abstract}

Pety et al. (2012) have reported the detection of eight transitions of a closed-shell, linear molecule (B11244) in observations toward the Horsehead PDR, which they attribute to the $l$-C$_3$H$^+$ cation.  Recent high-level \textit{ab initio} calculations have called this assignment into question; the anionic C$_3$H$^-$ molecule has been suggested as a more likely candidate.  Here, we examine observations of the Horsehead PDR, Sgr B2(N), TMC-1, and IRC+10216 in the context of both $l$-C$_3$H$^+$ and C$_3$H$^-$.   We find no observational evidence of $K_a = 1$ lines which should be present were the carrier indeed C$_3$H$^-$.  Additionally, we find a strong anti-correlation between the presence of known molecular anions and B11244 in these regions.  Finally, we discuss the formation and destruction chemistry of C$_3$H$^-$ in the context of the physical conditions in the regions.  Based on these results, we conclude there is little evidence to support the carrier is C$_3$H$^-$.

\vspace{1em}

\end{abstract}

\section{Introduction}
\label{intro}

Pety et al. (2012) have reported the detection of eight transitions of a closed-shell, linear molecule in observations toward the Horsehead photo-dissociation region (PDR).  They performed a spectroscopic analysis and fit to these transition frequencies and, based on comparison with the theoretical work (see Ikuta 1997 and refs. therein), attribute these transitions to the $l$-C$_3$H$^+$ cation.  Later, McGuire et al. (2013) identified the $J = 1 - 0$ and $J = 2 - 1$ transitions predicted by Pety et al. (2012) in absorption toward the Sgr B2(N) molecular cloud.  

$l$-C$_3$H$^+$ is important in the chemistry of hydrocarbons because reactions with this cation are thought to be the most important gas-phase channels to form other small hydrocarbons \citep{Turner2000,Wakelam2010}, like C$_3$H and C$_3$H$_2$, which are widely observed in different environments. However, the observed abundances of C$_3$H and C$_3$H$_2$ in PDRs are much higher than what pure gas-phase models predict. One possible explanation is that polycyclic aromatic hydrocarbons (PAHs) are fragmented into these small hydrocarbons in PDRs due to the strong UV fields (see, e.g., Fuente et al. 2003; Teyssier et al. 2004; Pety et al. 2005). The discovery of $l$-C$_3$H$^+$ thus brings further constraints to the formation pathways of the small hydrocarbons in these environments.

The attribution of these signals to the $l$-C$_3$H$^+$ cation, however, has since been disputed by Huang et al. (2013) and Fortenberry et al. (2013), with the latter suggesting the anion, C$_3$H$^-$, as a more probable carrier based on high-level theoretical work.  Because of the open question of identity, McGuire et al. (2013) use the convention of referring to the carrier as B11244, which we adopt here.

In this paper, we re-examine the observations of Pety et al. (2012) towards the Horsehead PDR, as well as \textbf{PR}ebiotic \textbf{I}nterstellar \textbf{MO}lecular \textbf{S}urvey (PRIMOS) observations of Sgr B2(N), the Kaifu et al. (2004) survey of TMC-1, and the Barry E. Turner Legacy survey of IRC+10216 in the context of discussing: ``What if B11244 is actually C$_3$H$^-$?"  In \S\ref{spectroscopy}, we discuss the spectroscopy of C$_3$H$^-$ using the properties derived by Fortenberry et al. (2013) and present simulated spectra.  In \S\ref{obs}, we briefly outline the observations used, and in \S\ref{analysis} discuss the analysis of these observations.  Finally, in \S\ref{results} we present the results of our findings and discuss them in the context of determining the identity of B11244.

\section{Spectroscopic Analysis}
\label{spectroscopy}

\begin{figure}
\includegraphics[scale=0.5]{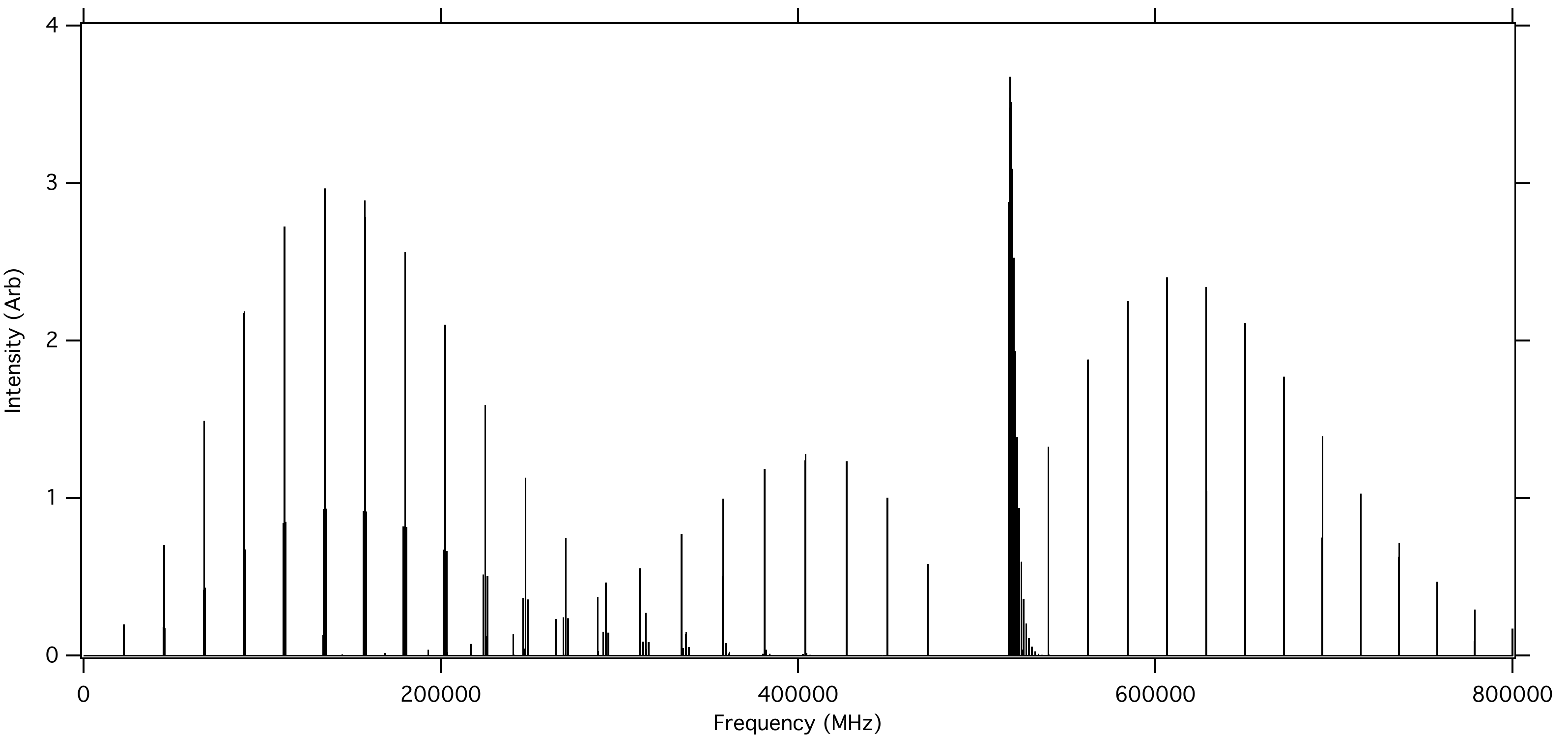}
\caption{Simulated spectrum of C$_3$H$^-$ at LTE, with an excitation temperature of $T_{ex} = 22$ K.}
\label{simulation}
\end{figure}

Table \ref{constants} provides the rotational constants and dipole moments used in this work to describe B11244, assuming it is $l$-C$_3$H$^+$ or C$_3$H$^-$.  The spectroscopic constants and fit for $l$-C$_3$H$^+$ are provided in Pety et al. (2012) and their predictive power confirmed in McGuire et al. (2013).  Fortenberry et al. (2013) provide a high-accuracy equilibrium structure for C$_3$H$^-$, rotational constants, and dipole moments.  These moments are not in the principal axis (PA) system, but can readily be converted to the PA system with a simple coordinate rotation resulting in $\mu_x \rightarrow \mu_a = 1.63$ Debye and $\mu_y \rightarrow \mu_b = 1.41$ Debye.  Indeed, the magnitude of this rotation is small, such that the value of these dipole moment remain essentially unchanged.  Assuming B11244 is C$_3$H$^-$, the observed transitions in Pety et al. (2012) and McGuire et al. (2013) are $a$-type, $K_a = 0$ transitions and thus ($B+C$) and $D_J$ can be well-determined from these lines.  To obtain these constants, the observed transitions from Pety et al. (2012) and McGuire et al. (2013) were fit using the CALPGM suite of programs.  An asymmetric-top Hamiltonian with a Watson S reduction in the I$^r$ representation was used.\footnote{Full details on the expressions and algorithms can be found in the CALPGM documentation and refs. therein. The interested reader may find that the analytical analysis presented by Polo (1957) provides a useful (and more approachable) approximation.}  The remaining constants were necessarily used as-is from the theoretical calculations.  A simulated spectrum of C$_3$H$^-$ at 22 K using this combined set of constants is displayed in Figure \ref{simulation}.  A full CALPGM catalog for C$_3$H$^-$ to 2 THz is also provided as Supplemental Information (see Table \ref{catalog}).

\begin{deluxetable}{l l l }
\tablecolumns{3}
\tabletypesize{\footnotesize}
\tablecaption{CALPGM catalog simulation of C$_3$H$^-$ format.}
\tablewidth{0pt}
\tablehead{
	\colhead{Column}	&	\colhead{Format}	&	\colhead{Description}
}
\startdata
			1		&	F13.4			&	Frequency (MHz)						\\
			2		&	F8.4				&	Error of Freq (MHz)						\\
			3		&	F8.4				&	Base 10 log intensity	 (nm$^2$MHz at 300 K)	\\
			4		&	I2				&	Degrees of freedom in partition function		\\
			5		&	F10.4			&	Lower state energy (cm$^{-1}$)				\\
			6		&	I3				&	Upper state degeneracy					\\
			7		&	I7				&	Species tag							\\
			8		&	I4				&	Quantum number format identifier			\\
			9		&	6I2				&	Upper state quantum numbers				\\
			10		&	6I2				&	Lower state quantum numbers				\\
\enddata
\tablecomments{For a complete description of this file format, see CALPGM documentation located at spec.jpl.nasa.gov.}
\label{catalog}
\end{deluxetable}

\begin{deluxetable}{l r c r c }
\tablecolumns{5}
\tabletypesize{\footnotesize}
\tablecaption{Rotational constants and dipole moments for B11244, assuming it is $l$-C$_3$H$^+$ or C$_3$H$^-$.}
\tablewidth{0pt}
\tablehead{
	\colhead{Constant}	&	\colhead{$l$-C$_3$H$^+$}	&	\colhead{Ref.}	&	\colhead{C$_3$H$^-$}	&	\colhead{Ref.}
}
\startdata
	$A$	(MHz)		&			\nodata			&	\nodata		&			529 134			&		(3)		\\
	$B$	(MHz)		&		11 244.9512(15)		&		(2)		&			11 355.3			&		(1)		\\
	$C$	(MHz)		&			\nodata			&	\nodata		&			11 134.5			&		(1)		\\
	\\
	$D$ (kHz)			&			7.766(40)			&		(2)		&		\nodata				&	\nodata		\\	
	$D_J$ (kHz)		&			\nodata			&	\nodata		&			4.63				&		(1)		\\
	$D_{JK}$ (kHz)		&			\nodata			&	\nodata		&			702				&		(3)		\\
	$D_K$ (MHz)		&			\nodata			&	\nodata		&			218				&		(3)		\\
	$d_1$ (Hz)		&			\nodata			&	\nodata		&			-112				&		(3)		\\
	$d_2$ (Hz)		&			\nodata			&	\nodata		&			-23				&		(3)		\\
	$H$ (Hz)			&			0.56(19)			&		(2)		&		\nodata				&	\nodata		\\	
	\\
	$\mu_a$ (Debye)	&			\nodata			&	\nodata		&			1.63				&		(1,3)		\\
	$\mu_b$ (Debye) 	&				3			&		(2)		&			1.41				&		(1,3)		\\				
\enddata
\tablerefs{(1) This work; (2) Pety et al. (2012); (3) Fortenberry et al. (2013).}
\label{constants}
\end{deluxetable}

\section{Observations}
\label{obs}

\textit{Sgr B2(N)} - The data presented toward Sgr B2(N) were obtained as part of the PRIMOS project using the Robert C. Byrd 100 m Green Bank Telescope.  The observed position was at (J2000) $\alpha =  $ 17$^{\mbox{h}}$47$^{\mbox{m}}$19$^{\mbox{s}}$.8, $\delta= -28^{\circ}22\arcmin17\arcsec$. An LSR source velocity of +64 km s$^{-1}$ was assumed.  Full observational details, including data reduction procedures and analysis, are given in Neill et al. (2012a).\footnote{Access to the entire PRIMOS dataset, specifics on the observing strategy, and overall frequency coverage information is available at http://www.cv.nrao.edu/$\sim$aremijan/PRIMOS/.}

\textit{IRC+10216} - The observations presented toward IRC+10216 are part of the Barry E. Turner Legacy Survey using the NRAO 12 m Telescope on Kitt Peak.  The observed position was at (J2000) $\alpha =  $ 9$^{\mbox{h}}$47$^{\mbox{m}}$57$^{\mbox{s}}$.3, $\delta= +13^{\circ}16\arcmin43\arcsec$.  An LSR source velocity of -26 km s$^{-1}$ was assumed.  Full observational details are given in Remijan et al. (2008).\footnote{All observations from the PRIMOS project and Barry E. Turner Legacy Survey are accessible at http://www.cv.nrao.edu/$\sim$aremijan/SLiSE.}

\textit{TMC-1} - The observations presented toward the TMC-1 dark cloud were taken as part of the Kaifu et al. (2004) survey using the Nobeyama Radio Observatory 45 m telescope.  The observed position was at (J2000) $\alpha = $ 4$^{\mbox{h}}$41$^{\mbox{m}}$42$^{\mbox{s}}$.5, $\delta= +25^{\circ}41\arcmin26.9\arcsec$.  An LSR source velocity of +5.85 km s$^{-1}$ was assumed.  Full observational details are given in Kaifu et al. (2004).

\textit{Horsehead PDR} - The observations presented toward the Horsehead PDR were taken with the IRAM 30-m telescope as part of the Horsehead WHISPER project (PI: J. Pety). The observed position was at (J2000) $\alpha = $ 5$^{\mbox{h}}$40$^{\mbox{m}}$53$^{\mbox{s}}$.936, $\delta= -2^{\circ}28\arcmin00\arcsec$.  An LSR source velocity of +10.7 km s$^{-1}$ was assumed.  Full observational details are given in Pety et al. (2012).

\section{Data Analysis}
\label{analysis}

The column density of B11244 in each source, assuming local thermodynamic equilibrium (LTE), can be calculated using Equation \ref{emissioncd} following the convention of Hollis et al. (2004).

\begin{equation}
\label{emissioncd}
N_T=\frac{3k}{8\pi^3}\times\frac{Q_re^{E_u/T_{ex}}}{\nu S\mu ^2}\times\frac{\sqrt{\pi}}{2ln2}\times\frac{\Delta T_A^* \Delta V/\eta _b}{1-\frac{(e^{(4.8\times10^{-5})\nu /T_{ex}}-1)}{(e^{(4.8\times10^{-5})\nu /T_{bg}}-1)}}\mbox{ cm}^{-2}
\end{equation}

Here, $N_T$ is the total column density, $Q_r$ is the rotational partition function, $E_u$ is the upper state energy in Kelvin, $T_{ex}$ is the excitation temperature, $\nu$ is the frequency of the transition in MHz, $S\mu^2$ is the transition strength in Debye$^2$, $\Delta T_A^*$ is the peak line intensity in mK, $\Delta V$ is the line FWHM in km s$^{-1}$, $\eta _b$ is the beam efficiency at frequency $\nu$, and $T_{bg}$ is the background temperature in Kelvin.

In the case of $l$-C$_3$H$^+$, the partition function is well approximated by the standard linear-molecule formula given by Eq. \ref{linearq} with $B$ expressed in Hz.  For C$_3$H$^-$, Eq. \ref{asymq}, with $\sigma = 1$ and rotational constants with units of MHz, is appropriate \citep{Gordy1984}. The accuracy of $Q_r$ for the anion is dependent on the accuracy of the rotational constants used. Thus, there is likely an uncertainty of a few percent in the value of $Q_r$ used here.  In any case, the partition function for the anion rapidly outpaces that of the cation above $T_{ex} \sim 8$ K.

\begin{equation}
Q_r \textup{ ($l$-C$_3$H$^+$)} \approxeq \frac{kT}{hB} = 1.85(T)
\label{linearq}
\end{equation}

\begin{equation}
Q_r \textup{ (C$_3$H$^-$)} \approxeq \frac{5.34\times 10^6}{\sigma}\left(\frac{T^3}{ABC}\right)^{1/2} = 0.65(T_{ex})^{3/2}
\label{asymq}
\end{equation}

To calculate upper limits of $l$-C$_3$H$^+$ in IRC+10216, we use the molecule-specific parameters given in McGuire et al. (2013) and the upper limit $\Delta T_A^*$ and $\Delta V$ values given in Table \ref{parameters}.  The line parameters and molecule-specific parameters used for all C$_3$H$^-$ calculations are given in Table \ref{parameters}.

\begin{deluxetable}{c c c c c c c c c c }
\tablecolumns{10}
\tabletypesize{\footnotesize}
\tablecaption{Observed and targeted transitions of B11244, assuming it is C$_3$H$^-$.  For simplicity, only those $K_a = 1$ transitions specifically searched for in our study are displayed.}
\tablewidth{0pt}
\tablehead{
\colhead{Transition}										&	\colhead{$\nu$}		&	\colhead{}							&	\colhead{E$_u$}		&	\multicolumn{2}{c}{Horsehead}					&	\multicolumn{2}{c}{Sgr B2(N)}						&	\multicolumn{2}{c}{IRC+10216}\\
\colhead{$J_{K_a,K_c}^{\prime} - J_{K_a,K_c}^{\prime\prime}$}	&	\colhead{(MHz)}		&	\colhead{S$_{ij}$}		& 	\colhead{(K)}			&	\colhead{$\Delta T_{mb}$}	&	\colhead{$\Delta V$}	&	\colhead{$\Delta T_A^*$}	&	\colhead{$\Delta V$}		&	\colhead{$\Delta T_A^*$}	&	\colhead{$\Delta V$} 
}	
\startdata
$1_{0,1} \rightarrow 0_{0,0}$								&	22 489.86				&	1				&	1.079				&	\nodata				&	\nodata			&	-27					&	13.4					&	\nodata				&	\nodata	\\
\\
$2_{1,2} \rightarrow 1_{1,1}$								&	44 755.94				&	1.5				&	28.07				&	\nodata				&	\nodata			&	\tablenotemark{a}		&	14.7					&	\nodata				&	\nodata	\\
$2_{0,2} \rightarrow 1_{0,1}$								&	44 979.50				&	2				&	3.237				&	\nodata				&	\nodata			&	-70					&	14.7					&	\nodata				&	\nodata	\\
$2_{1,1} \rightarrow 1_{1,0}$								&	45 197.57				&	1.5				&	28.10				&	\nodata				&	\nodata			&	$\leq 9$				&	14.7					&	\nodata				&	\nodata	\\
\\
$4_{1,4} \rightarrow 3_{1,3}$								&	89 510.82				&	3.75				&	35.59				&	$\leq 5.7$				&	0.81				&	\nodata				&	\nodata				&	\nodata				&	\nodata	\\
$4_{0,4} \rightarrow 3_{0,3}$								&	89 957.63				&	4				&	10.79				&	89					&	0.81				&	\nodata				&	\nodata				&	\nodata				&	\nodata	\\
$4_{1,3} \rightarrow 3_{1,2}$								&	90 394.13				&	3.75				&	35.70				&	$\leq 5.3$				&	0.81				&	\nodata				&	\nodata				&	\nodata				&	\nodata	\\
\\
$5_{1,5} \rightarrow 4_{1,4}$								&	111 887.54			&	4.8				&	40.96				&	$\leq 10.5$				&	0.81				&	\nodata				&	\nodata				&	\nodata				&	\nodata	\\
$5_{0,5} \rightarrow 4_{0,4}$								&	112 445.57			&	5				&	16.19				&	115					&	0.81				&	\nodata				&	\nodata				&	\nodata				&	\nodata	\\
$5_{1,4} \rightarrow 4_{1,3}$								&	112 991.72			&	4.8				&	41.11				&	$\leq 14.1$				&	0.81				&	\nodata				&	\nodata				&	\nodata				&	\nodata	\\
\\
$6_{0,6} \rightarrow 5_{0,5}$								&	134 932.69			&	6				&	22.66				&	72					&	0.81				&	$\leq$52				&	13					&	$\leq$5				&	27.5	
\\
\\
$7_{0,7} \rightarrow 6_{0,6}$								&	157 418.71			&	7				&	30.22				&	75					&	0.81				&	99\tablenotemark{b}		&	13					&	$\leq$7				&	27.5	\\
\\
$9_{0,9} \rightarrow 8_{0,8}$								&	202 386.75			&	9				&	48.56				&	62					&	0.81				&	\nodata				&	\nodata				&	\nodata				&	\nodata	\\
\\
$10_{0,10} \rightarrow 9_{0,9}$							&	224 868.40			&	10				&	59.36				&	30					&	0.81				&	\nodata				&	\nodata				&	\nodata				&	\nodata	\\
\\
$11_{0,11} \rightarrow 10_{0,10}$							&	247 348.23			&	11				&	71.23				&	45					&	0.81				&	\nodata				&	\nodata				&	\nodata				&	\nodata	\\
\\
$12_{0,12} \rightarrow 11_{0,11}$							&	269 826.05			&	12				&	84.18				&	25					&	0.81				&	\nodata				&	\nodata				&	\nodata				&	\nodata	\\
\enddata
\tablecomments{$\Delta T_A^*$ and $\Delta T_{mb}$ given in units of mK, $\Delta V$ given in units of km s$^{-1}$.  All upper limits are 1$\sigma$.  Values for the Horsehead PDR and Sgr B2(N) are based on Gaussian fits to the lineshapes.  The FWHM for IRC+10216 is based on a zeroeth-order approximation from other observed transitions.}
\tablenotetext{a}{Completely obscured by blends.}
\tablenotetext{b}{Partially blended.}
\label{parameters}
\end{deluxetable}	

While the observed $K_a = 0$ transitions allow us to constrain $B$ and $C$ reasonably well, the lack of any confirmed detection of a $K_a = 1$ transition limits the overall accuracy in predicting the frequencies of these lines.  However, the expected intensity of these lines, given a derived column density and temperature, is likely to be fairly accurate under LTE conditions.  In the Horsehead PDR, these lines should have a peak intensity of $T_{mb} \sim 20-28$ mK for $J^{\prime\prime} = 3$ to 6, using the derived conditions from the $K_a = 0$ transitions.  In Sgr B2(N), the expected intensities are below detectable values in our observations.

\begin{figure}
\plotone{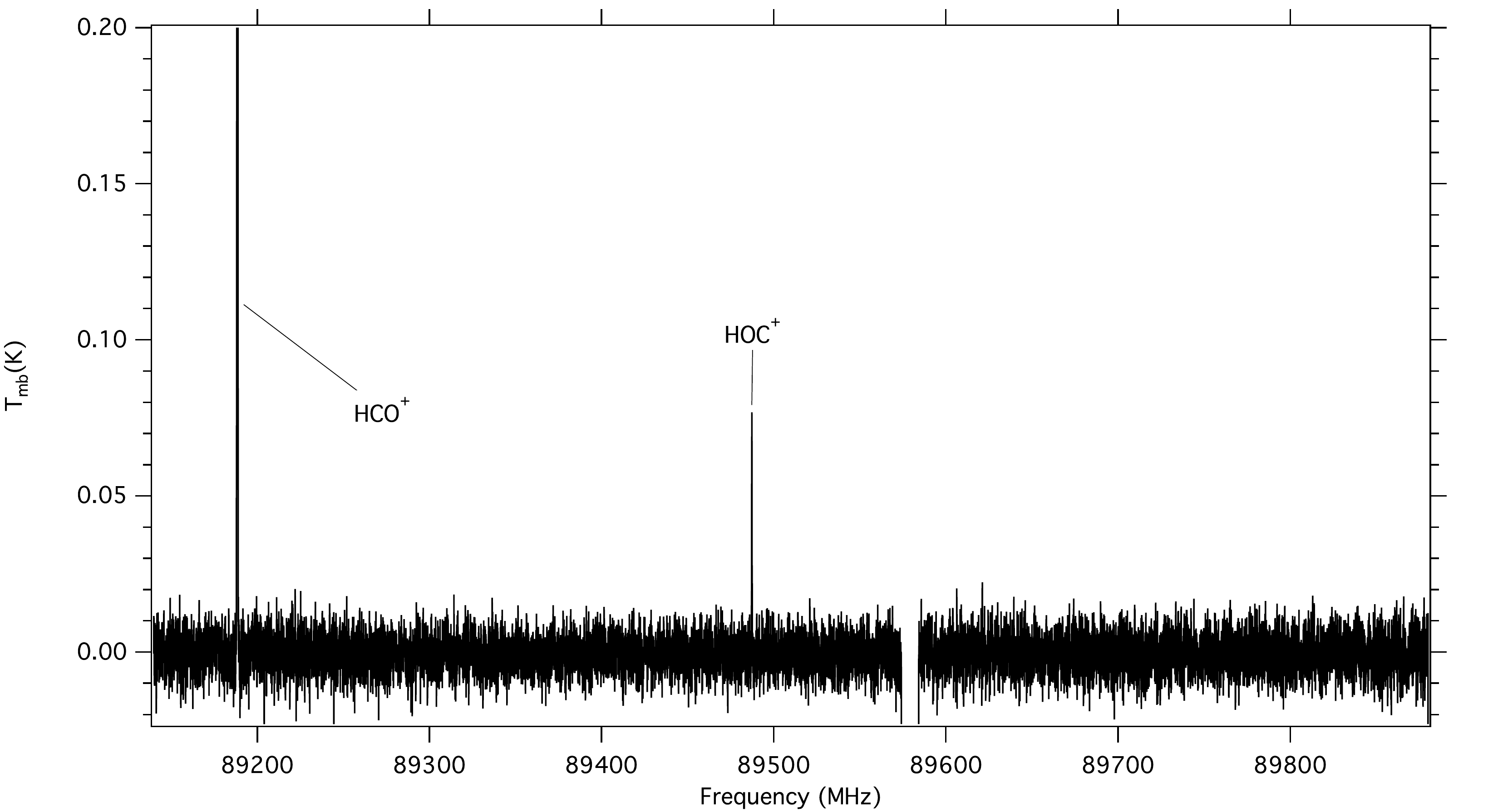}
\caption{Targeted frequency window around the predicted $K_a = 1$, $4_{1,4} - 3_{1,3}$ transition of C$_3$H$^-$ centered at 89535 MHz.  The RMS noise level is 5.8 mK.  Three features are observed - one each attributed to HCO$^+$ and HOC$^+$.  A third, located at $\sim$89580 MHz, has been positively identified as belonging to a known interstellar species, but has been removed from the spectra for proprietary reasons.  The identity of this line will be published in a forthcoming paper from Guzm\'{a}n et al.}
\label{satellites}
\end{figure}

We calculate a theoretical uncertainty in the center frequencies for these transitions of $\sigma\sim370$ MHz for the $J = 4 - 3$ transition to as much as $\sigma\sim650$ MHz for the $J = 7-6$ transition.  At LTE, the strongest of these lines fall within the 3 mm window of the Pety et al. (2012) survey.   Due to the uncertainties in the line centers, we have searched a region equal to each transition's uncertainty on either side of each predicted line center.  After identifying all known lines within this range, we find no detection of any signals which could be assigned to a $K_a = 1$ transition of C$_3$H$^-$ at the RMS noise level of the observations ($\sim$$5-10$ mK), despite peak predicted intensities of 20 - 28 mK at LTE.  An example spectrum of the region searched around the predicted $4_{1,4} - 3_{1,3}$ transition is shown in Figure \ref{satellites}.

At higher frequencies ($\nu > 500$ GHz), additional branches of C$_3$H$^-$ transitions are predicted, with slightly greater intensity.  We have no spectral coverage at these frequencies.  Additionally, these transitions are strongly dependent on the derived value for $A$, however, making any attempted search quite challenging.  

\section{Results \& Discussion}
\label{results}

In the following paragraphs, we discuss the results of our analysis in the context of determining the identity of B11244.  We do not address topics which have been previously covered in the literature, and for which our analysis provides no further information.  Namely, the agreement (or lack thereof) between the fitted rotational and distortion constants for each species with those calculated by Huang et al. (2013) and Fortenberry et al. (2013).

\subsection{Anion/neutral abundance ratio}

The results of fits to column density and excitation temperature in the Horsehead PDR and Sgr B2(N), and upper limits in IRC+10216 and TMC-1, are displayed in Table \ref{ratios} for $l$-C$_3$H$^+$, C$_3$H$^-$, and C$_6$H$^-$ as well as neutral C$_3$H and  C$_6$H.  In the Horsehead PDR and Sgr B2(N), the calculated column density for C$_3$H$^-$ is $\sim$3 times that of the cation.  This is due to an increase in the partition function and a decrease in the value of $S_{ij}\mu ^2$ for the anion.

Among the reported carbon-chain anionic species detected to date in the interstellar medium (ISM) (C$_4$H$^-$, C$_6$H$^-$, C$_8$H$^-$), C$_6$H$^-$ has been the most widely detected and characterized \citep{Gupta2009,Cordiner2013}.  The abundance fraction of C$_6$H$^-$,  relative to the neutral, is remarkably consistent across observed sources, varying from $\sim$$1.4\%$ to $4.4$\% \citep{Cordiner2013,McCarthy2006}.  The abundance ratio of C$_3$H$^-$ to neutral C$_3$H, which is more than an order of magnitude greater than that of C$_6$H$^-$ in observed sources, is therefore somewhat surprising.  Additionally puzzling is that C$_3$H$^-$ would appear to break the observed trend of increasing anion abundance fraction with increasing size, as well as the apparent trend for even-carbon molecular anions.  

Fortenberry et al. (2013) propose the most likely route to efficient formation of C$_3$H$^-$ is through a radiative attachment (RA) mechanism.  Herbst \& Osamura (2008) calculate an exceptionally low radiative attachment rate for C$_3$H. At 300 K, they find an attachment rate for C$_3$H orders of magnitude lower than for C$_4$H, C$_6$H, and C$_8$H.  Despite this, if B11244 is indeed C$_3$H$^-$, it would be the highest anion/neutral ratio detected in the ISM.\footnote{Cernicharo et al. 2008 find an abundance ratio of C$_5$N$^-$/C$_5$N in IRC+10216 of 57\%, but suggest it may in fact be as low as 12.5\%.}

As described by Fortenberry et al. (2013), C$_3$H$^-$ possesses both dipole-bound and valence excited states of the same multiplicity, which provide the necessary states to allow for a RA mechanism to form the anion \citep{Guthe2001,Carelli2013}.  Because the other detected anions possess only a dipole-bound state, Fortenberry et al. (2013) propose that the presence of the valence excited state may cause an enhancement in the production of C$_3$H. The extent of this enhancement is difficult to quantify, and thus we cannot say whether this can offset the lower RA rate predicted by Herbst \& Osamura (2008).

\begin{deluxetable}{l c c c c c c c c }
\tablecolumns{9}
\tabletypesize{\scriptsize}
\tablecaption{Column densities and excitation temperatures for $l-$C$_3$H$^+$ and C$_3$H$^-$ in our observations and from the literature, as well as ratios of these to their neutral counterparts.  Literature values for the ratio of C$_6$H$^-$ to neutral C$_6$H are also shown.}
\tablewidth{0pt}	
\tablehead{	\colhead{}			&	\multicolumn{2}{c}{$l$-C$_3$H$^+$}										&	\multicolumn{2}{c}{C$_3$H$^-$}								&	\colhead{C$_3$H}					&	$l$-C$_3$H$^+$/C$_3$H		&	C$_3$H$^-$/C$_3$H	&	C$_6$H$^-$/C$_6$H	\\
			\colhead{Source}	&	\colhead{$N$(10$^{11}$ cm$^{-1}$)}	&	\colhead{T$_{ex}$(K)}				&	\colhead{$N$(10$^{11}$ cm$^{-1}$)}	&	\colhead{T$_{ex}$(K)}	&	\colhead{$N$(10$^{11}$ cm$^{-1}$)}	&	\colhead{(\%)}					&	\colhead{(\%)}			&	\colhead{(\%)}			
}
\startdata
Horsehead PDR				&	4.8(9)$^{(1)}$						&	14(2)$^{(1)}$						&	12(1)\tablenotemark{a}				&	22(4)\tablenotemark{a}	&	21(7)$^{(1)}$						&	23(3)						&	57(16)				&	$\leq$9				\\
\vspace{-0.5em}\\
Sgr B2(N)						&	240(30)\tablenotemark{b}				&	8\tablenotemark{b}					&	790(90)\tablenotemark{b}				&	8\tablenotemark{b}		&	3000	(300)\tablenotemark{b}			&	8(1)							&	26(8)				&	\nodata\tablenotemark{c}	\\
\vspace{-0.5em}\\
IRC+10216					&	$\leq$6							&	32								&	$\leq$40							&	32					&	560								&	$\leq$1.1						&	$\leq$7.1				&	3(2)$^{(2)}$					\\
\vspace{-0.5em}\\
TMC-1						&	$\leq$6\tablenotemark{d}				&	9								&	$\leq$250\tablenotemark{d}			&	9					&	90$^{(3)}$								&	$\leq$7						&	$\leq$278				&	2.5(0.4)$^{(4)}$				\\
\enddata
\label{ratios}
\tablecomments{Uncertainties are given in parentheses in units of the last significant digit and are 1$\sigma$. All values were calculated for this work unless otherwise noted.}
\tablerefs{(1) Pety et al. (2012); (2) McCarthy et al. (2006); (3) Kaifu et al. (2004); (4) Cordiner et al. (2013)}
\tablenotetext{a}{To ensure consistency with the $l$-C$_3$H$^+$ values determined by Pety et al. (2012), these values have been determined via a rotation diagram analysis.  A least-squares fit analysis suggests this column density may actually be a factor of 2 higher.}
\tablenotetext{b}{These values are slightly revised from those in McGuire et al. (2013).  While re-examining the data for this study, it became clear that the transition at 45 GHz, regardless of the carrier, is likely highly sub-thermal.  We therefore base our figures here on only the 22.5 GHz transition, and assume the ``standard" 8 K excitation temperatures for cold molecules in this source.  We have extended this temperature to our previous analysis of C$_3$H in this source as well.}
\tablenotetext{c}{There have been no reported detections of, and we see no evidence for, the presence of C$_6$H$^-$  in Sgr B2(N).}
\tablenotetext{d}{Based on a tentative detection of only the 45 GHz transition which, like H$_2$CO, displays anomalous absorption against the 2.7 K CMB in this source.}
\end{deluxetable}

\subsection{Detection in Sgr B2(N)}

To our knowledge, no molecular anions have been detected in Sgr B2(N).  An examination of both the PRIMOS cm-wave data and the 2 mm Turner Survey shows no indication of the presence of any of the known molecular anions.  Of note, no such anions have been detected in the Horsehead PDR, either \citep{Agundez2008}.

\begin{figure}
\plotone{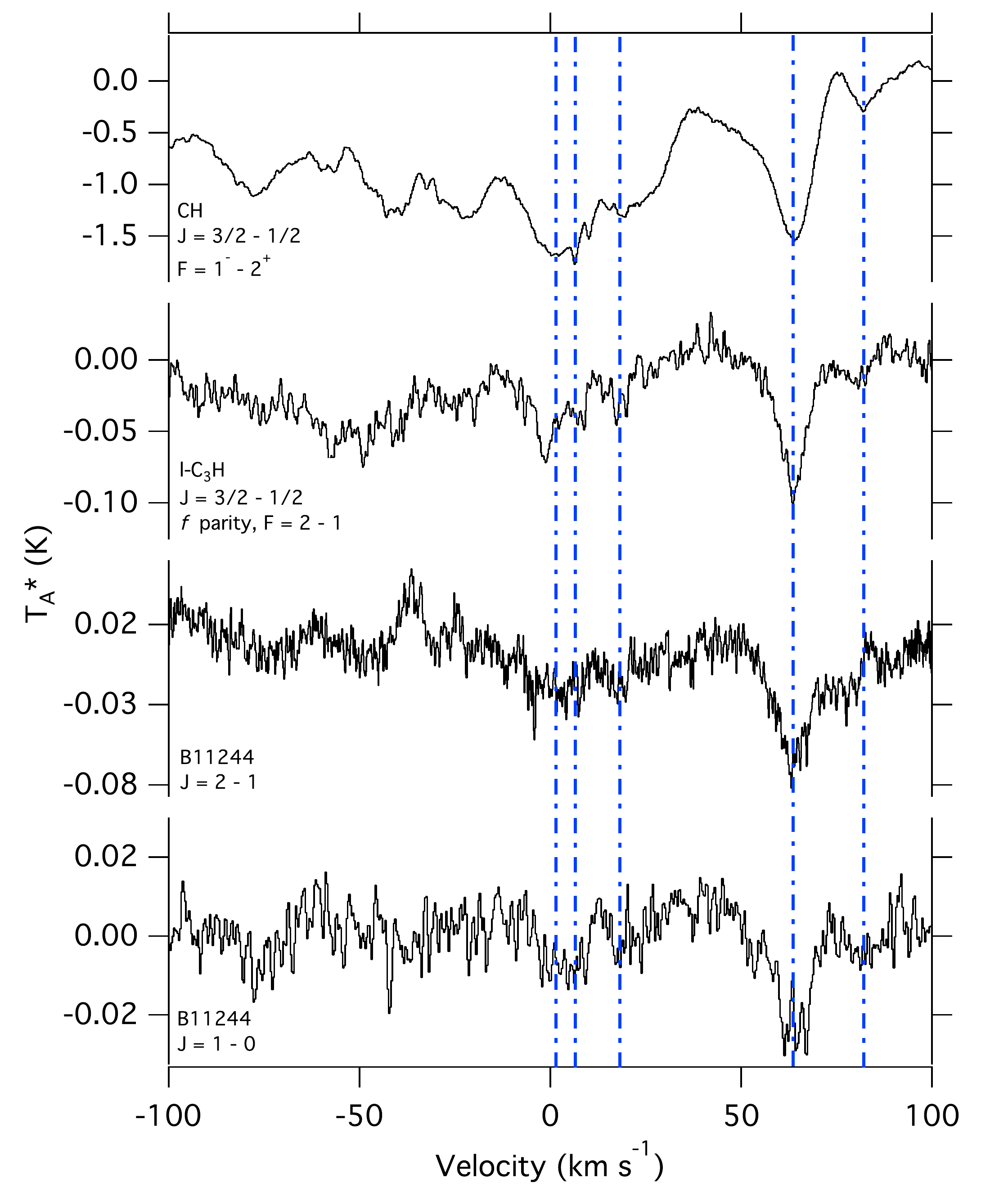}
\caption{Observed transitions of B11244 and $l$-C$_3$H toward Sgr B2(N) from PRIMOS and CH toward Sgr B2(N) from HEXOS.  The blue vertical lines are provided to guide the eye to the diffuse cloud velocity components.  The velocity axis is referenced to the rest frequency of each transition.}
\label{spiralarms}
\end{figure}

However, a re-examination of the PRIMOS data originally presented in McGuire et al. (2013) finds some evidence for B11244 absorption in lower-velocity ($V_{LSR} \sim +0 - 10$ km s$^{-1}$ and $V_{LSR}  \sim +18$ km s$^{-1}$) diffuse clouds along the line of sight to Sgr B2(N).  For illustration, the $J = 1-0$ and $J = 2 - 1$ transitions of B11244 are shown in Figure \ref{spiralarms} in comparison to the known CH absorption spectra toward SgrB2(N)\footnote{The CH spectra shown are from the HEXOS survey of Sgr B2(N).  Full observational, reduction, and analysis details are available in Neill et al. (2012b) and Neill et al. (2013).}.  The strongest observed transition of $l$-C$_3$H is also shown and displays low-velocity absorption as well, although the $\sim$0 km s$^{-1}$ component is blended with the +64 km s$^{-1}$ main component of the $l$-C$_3$H, $J = 3/2 -1/2$, $f$-parity, $F = 1-0$ transition.   

While numerous cationic species have been detected in these diffuse clouds (see, e.g., Gerin et al. 2010 \& Godard et al. 2010), the possibility of anion chemistry in these regions is not well-understood.  Indeed, no anions have previously been seen in these line-of-sight clouds. The indication of B11244 in these regions, presented here, will hopefully be a motivating factor which will drive future studies.  Investigations of this diffuse gas, which displays chemistry distinct from regions such as IRC+10216 and Sgr B2(N), will certainly prove invaluable in furthering our understanding of gas-phase ion chemistry.

\subsection{Non-detection in IRC+10216}

IRC+10216 has been the preeminent source for the detection of anionic species.  The carbon-chain anions C$_4$H$^-$, C$_6$H$^-$, and C$_8$H$^-$  have all detected in this source \citep{Gupta2007,Cernicharo2007,McCarthy2006,Remijan2007}, as well as the cyano-anions CN$^-$, C$_3$N$^-$, and C$_5$N$^-$ \citep{Agundez2010,Thaddeus2008,Cernicharo2008}.  As shown in Table \ref{parameters}, however, we see no signal from B11244 toward this source at an RMS of $\sim$5-7 mK.  Given the known abundance of the neutral C$_3$H, we can determine upper limits to the anion fraction.  We assume a rotational temperature of 32 K - similar to that of C$_6$H$^-$ and C$_8$H$^-$ in this source, and slightly higher than that of C$_4$H$^-$.  This results in an upper limit abundance fraction of C$_3$H$^-$/C$_3$H of only $\sim$7\% (see Table \ref{ratios}), about twice that of C$_6$H$^-$ and less than half of the lower edge of our error bars in Sgr B2(N).

\subsection{Anion destruction via photodetachment}

Kumar et al. (2013) have recently shown that UV photodetachment may be the dominant destruction mechanism of interstellar anions in IRC+10216.  Their models assume the standard interstellar UV radiation field (c.f. Draine 1978), and find UV photodetachment is significant at these values.  In the Horsehead PDR, however, the UV field is $\sim$60 times the standard UV value \citep{Habart2005}.  It is therefore contradictory that despite a far higher (destructive) UV field, C$_3$H$^-$ would be present in the Horsehead PDR with an anion:neutral ratio 8 times higher than the upper limit for IRC+10216.

\subsection{Non-detection of $K_a = 1$ transitions}

The lack of detection of any signal which could reasonably be attributed to emission from $K_a = 1$ transitions strongly disfavors the assignment of B11244 to C$_3$H$^-$. In the Horsehead PDR, the $K_a = 0$ transitions of B11244 are well-modeled by LTE assumptions, and thus we do expect the intensity of the $K_a = 1$ transitions to be reasonably well predicted.  It should be noted, however, that a single molecule can display distinctly different excitation temperatures and column densities between two $K$ ladders.  Indeed, previous observations of HNCO toward OMC-1 show distinct differences between LTE column density and temperature measurements in high-$K$ and low-$K$ ladders \citep{Blake1987}.  The authors attribute this to radiative excitation of the higher-$K$, higher-energy states through strongly allowed b-type transitions at far-infrared (far-IR) wavelengths.

In the case of C$_3$H$^-$, we can examine three limiting cases which may apply in the Horsehead PDR: LTE conditions, the influence of a weak far-IR radiation field, and the influence of a strong far-IR radiation field.

\subsubsection{LTE}

As shown in \S\ref{analysis}, under LTE conditions the $K_a = 1$ transitions of C$_3$H$^-$ have predicted intensities of 20 - 28 mK.  These are clearly not detected in our spectra at the RMS noise level of the observations ($\sim5-10$ mK).  Thus we can say with some certainty that C$_3$H$^-$ is not present in the Horsehead PDR under LTE conditions.

\subsubsection{Weak far-IR radiation field}

In the case of a weak far-IR radiation field, and assuming low to moderate H$_2$ densities in the region (i.e. non-LTE), the population of the $K_a =$ 1, 2, ... levels will be largely dominated by radiative selection rules.  Any population driven into the $K_a  =$ 1, 2, ... levels by collisions will rapidly decay back into the $K_a = 0$ states via radiative emission.  For C$_3$H$^-$, this will result in a decrease in the observed intensity of the $K_a = 1$ transitions, relative to the $K_a = 0$ transitions, as compared to LTE.  In this case, the lack of detected $K_a = 1$ transitions does not provide a constraint on the presence of C$_3$H$^-$.

\subsubsection{Strong far-IR radiation field}

We now examine the case of a strong far-IR radiation field, and low to moderate H$_2$ densities in the region (i.e. non-LTE).  For transitions arising from low-energy states, the relative populations will be determined by the rotational excitation temperature of the molecule ($T_{ex}$ = 22 K).  For transitions from higher-energy states connected by far-IR transitions, the relative populations of the energy levels will be determined by the color temperature of the dust radiation field at that frequency if that temperature is higher than $T_{ex}$.  Thus, the $K_a =$ 1, 2, ... transitions would be relatively more intense than predicted by LTE simulations at $T_{ex}$.  This is the case for HNCO in OMC-1 \citep{Blake1987}, where the higher-$K$ transitions are more intense than predicted from observations of the lower-$K$ transitions.

In the Horsehead PDR, Goicoechea et al. (2009) measure the millimeter dust continuum to have a temperature $T_d \simeq 30$ K.  Under these circumstances, we would expect the $K_a = 1$ transitions to have intensities of $\sim$$27 - 34$ mK.  We can therefore conclude that assuming B11244 is subject to the radiation field measured by Goicoechea, then C$_3$H$^-$ is not the carrier.

It is clear from the two non-LTE cases discussed above that the location of B11244 within the Horsehead PDR region is critical.  Interferometric mapping of the location of B11244, relative to the measured continuum levels in this region, will provide considerable insight into the mechanisms at work.

\section{Conclusions}

We have presented an analysis of observations of the Horsehead PDR, Sgr B2(N), IRC+10216, and TMC-1 with the goal of determining the identity of B11244.  Our findings can be summarized as follows.

\begin{enumerate}

\item If B11244 is C$_3$H$^-$, it would display the highest anion:neutral ratio yet observed in the ISM (57\% in the Horsehead PDR).
\item We find no evidence for C$_3$H$^-$ emission in observations toward IRC+10216 and place an upper limit on the anion:neutral ratio in this source well below that found in the Horsehead PDR and Sgr B2(N).
\item Recent work has shown UV photodetachment is a dominant destruction pathway for molecular anions \citep{Kumar2013}.  Despite a UV field more than 60 times that of IRC+10216, C$_3$H$^-$ would be present in the Horsehead PDR with an anion:neutral ratio more than 8 times that of the upper limit in IRC+10216.
\item We find no evidence for the $K_a = 1$ lines of C$_3$H$^-$ in observations of the Horsehead PDR.  We examine three limiting cases for conditions within the Horsehead PDR and find that a weak far-IR radiation field can account for the lack of observed $K_a = 1$ transitions.  LTE conditions or the presence of a strong far-IR radiation field, however, strongly disfavor the presence of C$_3$H$^-$.  A significant far-IR radiation field has been reported for the Horsehead PDR, but it is unclear whether B11244 is subject to this radiation.

\end{enumerate}

The observational evidence presented here, taken as a whole, casts doubt on the assignment of B11244 to C$_3$H$^-$, favoring instead the cation, $l$-C$_3$H$^+$ as the most likely candidate.  The evidence is, however, circumstantial; a definitive answer will almost certainly require laboratory confirmation.  Indeed, K.N. Crabtree and co-workers at the Harvard-Smithsonian Center for Astrophysics have undertaken such work using Fourier-transform microwave spectroscopy.  Preliminary evidence is suggestive of the cationic species.  The full results of the laboratory investigation will be published in an upcoming paper (K.N. Crabtree, Private Communication).

Additional observations of the Horsehead PDR, with the aim of detecting the $b$-type transitions of C$_3$H$^-$, predicted to be strongest between 500 - 600 GHz, would also provide further evidence.  Perhaps more insightful would be interferometric observations to discover the spatial correlation, or lack thereof, of B11244 with the previously observed far-IR radiation field. Finally, further observations of the diffuse gas along the sightline to Sgr B2(N) would likely prove fruitful in understanding the possibility of anion chemistry in these regions. 

\acknowledgments

We are grateful to the anonymous referee for helpful comments which have greatly improved the quality of this manuscript.  We thank E.A. Bergin and J.T. Neill for providing the HEXOS spectra presented here, and M. Ohishi for providing the observational data toward TMC-1.  We also thank H. Gupta, M. Cordiner, and A. Faure for insightful conversations.  B.A.M. gratefully acknowledges funding by an NSF Graduate Research Fellowship.  V.G. thanks support from the Chilean Government through the Becas Chile scholarship program.  This work was partially funded by grant ANR-09-BLAN-0231-01 from the French Agence Nationale de la Recherche as part of the SCHISM project.  The National Radio Astronomy Observatory is a facility of the National Science Foundation operated under cooperative agreement by Associated Universities, Inc.

\end{document}